\documentclass[preprint2,twoside]{hwo}

\usepackage{graphicx}
\usepackage{enumitem}
\usepackage{tabularx}

\newcommand{\add}[1]{#1}

\input{hwo.h}

\begin{document}

\title{\textbf{\LARGE Direct imaging characterization of cool gaseous planets}}

\author {\textbf{\large Michiel Min$^{1}$}}
\affil{$^1$\small\it SRON, Space Research Organisation Netherlands, Niels Bohrweg 4, 2333 CA, Leiden, The Netherlands, \email{M.Min@sron.nl}}

\author {\textbf{Jo Barstow$^{2}$}}
\affil{$^2$\small\it School of Physical Sciences, The Open University, Walton Hall, Milton Keynes, MK7 6AA, UK}

\author {\textbf{Laura C. Mayorga$^{3}$}}
\affil{$^3$\small\it Johns Hopkins Applied Physics Laboratory, 11100 Johns Hopkins Rd, Laurel, MD 20723, USA}

\author {\textbf{Hannah Wakeford$^{4}$}}
\affil{$^4$\small\it University of Bristol, School of Physics, HH Wills Physics Laboratory, Tyndall Avenue, Bristol, UK}

\author {\textbf{Jason Wang$^{5}$}}
\affil{$^5$\small\it Center for Interdisciplinary Exploration and Research in Astrophysics (CIERA),
Northwestern University, 1800 Sherman Ave, Evanston, IL, 60201, USA}

\author {\textbf{Renyu Hu$^{6,7}$}}
\affil{$^6$\small\it Jet Propulsion Laboratory, California Institute of Technology, Pasadena, CA 91011, USA} 
\affil{$^7$\small\it Division of Geological and Planetary Sciences, California Institute of Technology, Pasadena, CA 91125, USA} 

\author{\footnotesize{\bf Contributing authors:}
Beth Biller (University of Edinburgh, UK), 
José A. Caballero (CSIC-INTA, Madrid, Spain), 
Ludmila Carone (Space Research Institute Graz, Austria), 
Sarah Casewell (University of Leicester, UK),
Katy L. Chubb (University of Bristol),
Mario Damiano (JPL),
Siddharth Gandhi (University of Warwick),
Antonio García Muñoz (CEA Paris-Saclay, France), 
Christiane Helling (Space Research Institute Graz, Austria), 
Finnegan Keller (Arizona State University),
Nataliea Lowson (University of Delaware),
Evert Nasedkin (Trinity College Dublin),
Ryan MacDonald (University of Michigan), 
Jean-Baptiste Ruffio (University of California), 
Evgenya Shkolnik (Arizona State University), 
Christopher C. Stark (NASA Goddard Space Flight Center)
}

\author{\footnotesize{\bf Endorsed by:}
Narsireddy Anugu (Georgia State University),
Reza Ashtari (JHU-APL),
Jayne Birkby (University of Oxford),
Alan Boss (Carnegie Science),
Adam Burgasser (UC San Diego),
Aarynn Carter (STScI),
Jessie Christiansen (Caltech/IPAC),
Nicolas	Crouzet (Kapteyn Astronomical Institute, University of Groningen, The Netherlands),
Darío González Picos (Leiden Observatory),
Caleb Harada (UC Berkeley),
Theodora Karalidi (University of Central Florida),
Preethi	Karpoor	(Indian Institute of Astrophysics),
James Kirk (Imperial College London),
Alen Kuriakose (KU Leuven, Belgium),
Adam Langeveld (Johns Hopkins University),
Eunjeong Lee (EisKosmos (CROASAEN), Inc.),
Stanimir Metchev (University of Western Ontario),
Eric Nielsen (Department of Astronomy, New Mexico State University),
William Roberson (New Mexico State University),
Farid Salama (NASA Ames Research Center)
Maissa Salama (University of California, Santa Cruz),
Aniket Sanghi (Caltech),
Edward Schwieterman (Department of Earth and Planetary Sciences, University of California, Riverside),
Tomas Stolker (Leiden University),
Peter Wheatley (University of Warwick, UK)
}



\begin{abstract}
Cool gas giant exoplanets, particularly those with properties similar to those of Jupiter and Saturn, remain poorly characterized due to current observational limitations. This white paper outlines the transformative science case for the Habitable Worlds Observatory (HWO) to directly image and spectroscopically characterize a broad range of gaseous exoplanets with effective temperatures below 400 K. The study focuses on determining key atmospheric properties, including molecular composition, cloud and haze characteristics, and temperature structure, across planets of varying sizes and orbital separations. Leveraging reflected light spectroscopy and polarimetry, HWO will enable comparative planetology of cool gas giants orbiting both solar-type and M-dwarf stars, bridging the observational gap between hot exoplanets and Solar System giants. We present observational requirements and survey strategies necessary to uncover correlations between atmospheric properties and planetary or stellar parameters. This effort will establish critical constraints on planetary formation, cloud microphysics, and the role of photochemistry under diverse irradiation conditions. The unique capabilities of HWO will make it the first facility capable of characterizing true exo-Jupiters in reflected light, thus offering an unprecedented opportunity to place the Solar System in a broader galactic context.
\\
\\
\end{abstract}

\vspace{2cm}

\section{Science Goal}

The key science question for this document is:
\begin{enumerate}[label=\textbf{$\star$}]
\item \emph{What are the properties of individual gas giant planets, and which processes lead to planetary diversity?}
\end{enumerate}

Giant planets are so far the best studied extrasolar planetary mass objects, but currently the types of giant planets we can study in detail are limited. Measurements that allow us to determine the composition of planetary atmospheres can currently either be performed for transiting planets, which generally orbit very close to their parent stars and are often highly irradiated; non-transiting, but close to edge-on, planets in tight orbits that have large radial velocity amplitudes; and massive, young, giant planets at wide separations from their parent stars that are amenable to direct imaging. Exoplanets at similar temperatures and orbital separations to the Solar System gas giants are inaccessible with current technology, representing a significant gap in our understanding of giant planet composition, dynamics and evolution. 

Over the coming decades various observational facilities will come online that have the ability to fill this observational void in some capacity. The Nancy Grace Roman CGI will allow direct detection of reflected light from Jupiter analogs around the nearest stars. Also, the coming decade will see the first ELT instruments come online that will be able to image and take spectra of a variety of gas giant planets in different orbital configurations. However, all these observatories are limited in either inner working angle (for the Roman CGI) or contrast with respect to the star (limited for the ELT by the atmosphere of the Earth). The Habitable Worlds Observatory provides the opportunity to map a much larger part of parameter space, giving access to planets of a variety of different sizes and temperatures. This will allow more detailed comparative planetology. We provide more details on the sample statistics expected with ELT and Roman CGI compared to HWO further on in this document.

Characterizing gas giant planets in reflected light at different distances from their parent stars, and around stars of different temperatures, will also provide us with an insight into cloud particle characterization. HWO will allow us to survey giant planets across this parameter space and investigate correlations between aerosol properties and the irradiation (starlight) environment of the planet. The formation of cloud particles is determined by the local thermodynamic conditions, while the formation of photochemical hazes (which can serve as cloud condensation nuclei) are triggered by an external radiation field. For a subset of planets we will be able to obtain time-resolved spectroscopy, providing additional insight into the homogeneity of cloud coverage.

In this document we focus on performing characterization and comparative planetology for a range of gaseous planets with masses between Neptune and Jupiter. The majority of the discussion will be focused on relatively cool (i.e. $<400$\,K) planets, but we might have access to warmer planets for some very nearby stars.

\subsection{What Fundamental Planetary Parameters and Processes Determine the Complexity of Giant Planet Atmospheres?}

We want to understand how the gas-phase composition, cloud and haze properties, and structure of gas giant atmospheres varies across different planetary environments and properties, including
\begin{itemize}
    \item Fundamental planetary properties (mass, radius and rotation rate)
    \item Irradiation properties (total irradiation, UV irradiation)
    \item Planetary system properties (other planets in the system)
\end{itemize}

We want to characterize a sufficiently large sample of planets to answer the following questions:
\begin{enumerate}[label=\textbf{$\star$}]
    \item \emph{Are planets with the same fundamental properties and irradiation environment similar?}
    \item \emph{What is the impact of different formation histories on exoplanet properties, and are these impacts quantifiable for mature planets?}
    \item \emph{How do aerosols (clouds and hazes) impact the energy budget of an atmosphere? How would the presence of aerosols impact the measured observations and how do those aerosols impact the planetary environment/climate?}
\end{enumerate}

\subsection{How Does a Planet’s Interaction with Its Host Star and Planetary System Influence Its Atmospheric Properties over All Time Scales?}

In conjunction with what we hope to learn for the previous question, we also want to learn how planetary properties vary as a function of time, and how this relates to the time-varying stellar environment. In this context the varying UV irradiation from the central star plays an important role. How does this affect the photochemistry in giant planet atmospheres?

\subsection{How Do Giant Planets Fit Within a Continuum of Our Understanding of All Substellar Objects?}

We seek to identify trends and patterns in planetary atmosphere properties to see how they relate to other planetary properties and/or factors of the planet’s environment. We also want to learn what are the fundamental differences and similarities between giant planets and brown dwarfs, to try to understand the impact of the differing formation scenarios for isolated brown dwarfs compared to giant planet companions.


\section{Science Objective}

\subsection{Determine the atmospheric characteristics of a wide variety of gas giant planets}

In order to answer the above-mentioned questions, we need to determine the characteristics of atmospheres of a significant sample of planets with varying: 
\begin{itemize}
    \item Mass / Radius → We aim at masses/radii of (sub) Neptune to a few times Jupiter
    \item Orbital separation → We aim at orbits representing the irradiation temperatures below ~400\,K
    \item Central star properties → We focus on Solar type stars and M-dwarfs
\end{itemize}
For these planets we need to determine:
\begin{itemize}
    \item The molecular composition
    \item The influence and general properties of clouds and hazes
    \item The atmospheric temperature structure
\end{itemize}
To determine weather systems on these planets we need to determine the cloud map of the planet and how it changes over time (time variability of the system). Rapidly-rotating, Jupiter analog planets may be observed repeatedly at relatively high cadence over a short timescale, which will also allow determination of the rotation period.  

\subsubsection{Cloud properties}

Reflected light spectroscopy of giant exoplanets is a key tool to determine cloud properties on these objects. However, so far this has largely been explored only theoretically, due to the challenges of obtaining reflected light spectra for transiting exoplanets \citep[e.g.][]{Evans_2013}. 
\add{Nevertheless, recent observations are beginning to provide empirical constraints. \citet{Fraine_2021} used broadband HST/WFC3 photometry to measure the secondary eclipse of WASP-43b, placing a stringent upper limit on its geometric albedo and demonstrating the potential of UVIS for probing reflected light from hot Jupiters. More recently, \citet{Radica_2025} presented HST/WFC3 UVIS spectroscopy of the ultra-hot Neptune LTT 9779b, setting a 3$\sigma$ upper limit of 113 ppm on the eclipse depth in the 0.2–0.8$\,\mu$m band and using forward modeling to show that silicate clouds likely explain the planet’s elevated reflectivity. Together, these studies demonstrate that even low-resolution reflected light measurements can constrain aerosol properties and cloud composition when combined with phase-resolved or multi-band observations.}

\add{In addition to gas giant planets, irradiated binary brown dwarfs might provide an additional interesting resource to investigate the effects of irradiation on cool atmospheres and potentially investigate brown dwarf formation processes through abundance analysis. There is now a sample of $\sim$50 transiting brown dwarf planets known \citep{Vowell_2025} and some information on their behavior in reflected light is unfolding. For example, the $TESS$ secondary eclipse depth of KELT-1b, a 27~M$_{\rm Jup}$ BD orbiting a 6500~K F star every 1.33 days, is deeper than expected. These results were interpreted as being caused by a high albedo caused by reflective high-altitude silicate clouds being blown onto the dayside from the cooler nightside \citep{Beatty_2020} or by collisionally induced absorption in the atmosphere \citep{vonEssen_2021}.  Shorter wavelength $CHEOPS$ photometry conversely suggested a brightness temperature significantly cooler than was predicted by $Spitzer$ phase curves, and inconsistent with the $TESS$ eclipse depths, interpreted by \citet{Parviainen_2023} as due to a variable albedo caused by cloud variability. These results highlight the need for broadband, time resolved observations to constrain cloud formation in these high gravity atmospheres.}

Models by \citet{Sudarsky_2000} show that the albedo spectra change dramatically with different condensing species across types of giant exoplanet from Jupiter analogs through to ultrahots, with their Class I (Jupiter-like, ammonia clouds) and Class II (water cloud dominated) spectra representing what we are likely to encounter in our proposed sample. \citet{Cahoy_2010} examine the effects of planetary metallicity and orbital distance on the reflected light spectrum of Jupiter- and Neptune-like planets, finding that albedo across the optical peaks for 2\,AU distances. Models by \citet{Hu_2019} find that the cloud top pressure can be recovered for cool Jupiters with methane-dominated spectra. 

\begin{figure*}[ht!]
    \centering
    \includegraphics[width=0.75\textwidth]{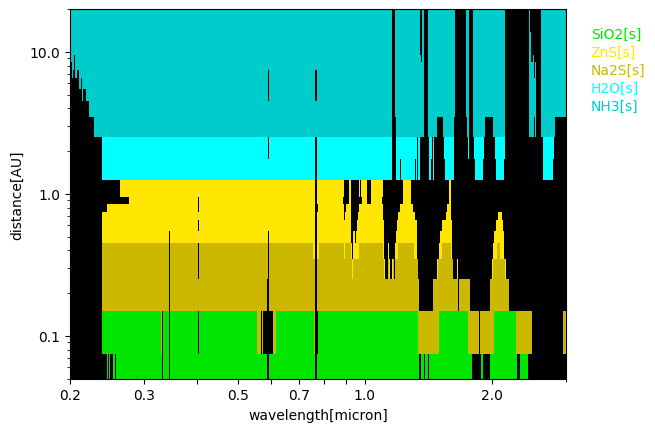}
    \caption{Cloud composition at the $\tau=1$ surface as a function of wavelength and separation from the central star. Black regions are where the atmosphere is dominated by molecular opacity.}
    \label{fig:cloudspecies}
\end{figure*}

Fig.~\ref{fig:cloudspecies} shows simulations of a Jupiter mass planet around a Solar type star, placed at varying separations from the star. The simulations include cloud formation and are iterated to be in radiative convective equilibrium \citep[computed with ARCiS version July 2025, see][]{Min2020, Chubb2022, Huang2024}. As a benchmark we compute Jupiter around the Sun to check if the simulations can predict what is known about Jupiter. The results of this comparison can be found in Appendix A. Fig.~\ref{fig:cloudspecies} shows the dominant cloud species above an optical depth of one (black is if there is no significant cloud column above the optical depth of one surface). It can be seen that clouds are best probed at visible wavelengths and between the molecular bands of methane and water in
the NIR. Also, it is clear that the types of clouds will vary with distance from the central star. Note that the simulations presented here are for a specific set of cloud formation parameters (nucleation rate and location, atmospheric diffusion…) and different choices here can lead to clouds higher/lower in the atmosphere.
The different cloud species we find have very different refractive indices allowing to discriminate between them especially with phase-resolved polarimetry. This has been done by e.g. \citet{Hansen1974} to show that the clouds of Venus are composed of sulfuric acid.

\begin{figure*}[ht!]
    \centering
    \includegraphics[width=0.75\textwidth]{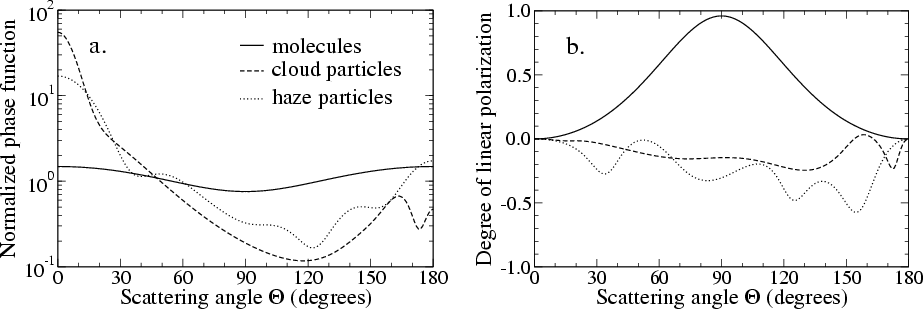}
    \includegraphics[width=0.75\textwidth]{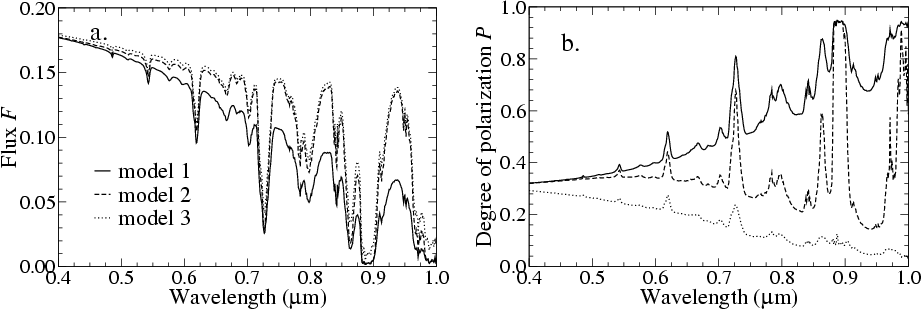}
    \caption{The typical scattering properties of molecules, cloud and haze particles (upper panels) and the simulated total intensity and polarized intensity spectra of a typical Jupiter type planet. Model 1 represents a clear atmosphere, model 2 has a tropospheric cloud layer and model 3 has a stratospheric haze layer. Figures taken from \citet{Stam2004}, used with permission.}
    \label{fig:Stam2004}
\end{figure*}

As can be seen in Fig.~\ref{fig:Stam2004} \citep[taken from][]{Stam2004}, the polarization and scattering properties of molecules and cloud/haze particles are very different. Where molecules create very high polarization at 90 degrees scattering and a very smooth curve, cloud particles create a lower polarization phase curve which shape is very dependent on the composition, size and shape of the cloud particles. Therefore, for very cloudy atmospheres the degree of polarization will be lower at the wavelengths where we probe the cloud layers (see the lower right panel of Fig.~\ref{fig:Stam2004}). If the cloud or haze layer is high up, it can even mute the polarization signature over the entire wavelength range.

\subsubsection{Molecular composition}

\begin{figure*}[ht!]
    \centering
    \includegraphics[width=0.75\textwidth]{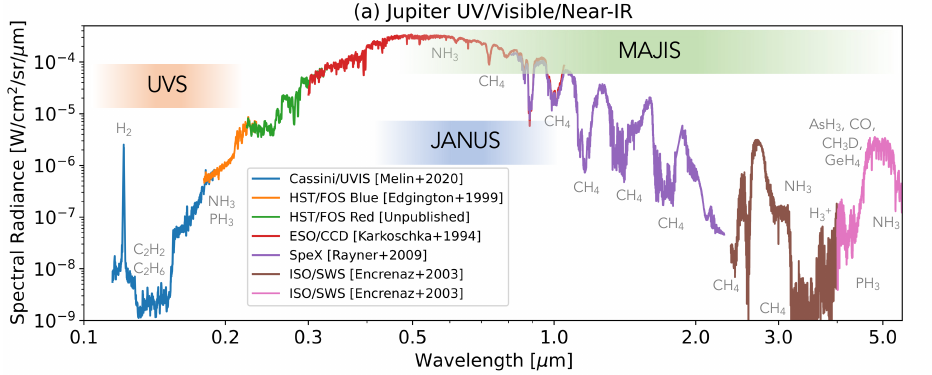}
    \caption{The observed UV-optical-near infrared spectrum of Jupiter, taken from \citet{Fletcher2023} “Jupiter Science Enabled by ESA’s Jupiter Icy Moons Explorer”, used with permission. Methane, ammonia, phosphine, acetylene and ethane are all accessible between 0.1 and 2.0 $\mu$m.}
    \label{fig:Fletcher2023}
\end{figure*}

Reflected light spectra of gas giant exoplanets contain the imprint of several molecular species. The observed UV-optical-near infrared spectrum of Jupiter (Fig.~\ref{fig:Fletcher2023}) highlights the presence of absorption by methane, ammonia, phosphine, acetylene, and ethane (shortwards of 2\,$\mu$m), and arsine, germine, carbon monoxide and deuterated methane at longer wavelengths. H$_3^+$ auroral emission is also observed longward of 3\,$\mu$m, and molecular hydrogen emission in the ultraviolet. Measuring reflected light absorption and emission signatures of these gases is key if we are to characterize cool and temperate Jupiters. In addition, for gas giant planets only a little warmer than Jupiter, \citet{MacDonald_2018} demonstrate that water vapor absorption should be readily detectable at optical and near-infrared wavelengths for lower metallicity planets in the absence of optically thick high altitude clouds.
\add{Disequilibrium chemistry is likely to be important in cool giant exoplanets, where slower chemical kinetics can preserve quenched abundances of species like CO and NH$_3$, depending on the strength of vertical mixing.}

Several examples exist of retrieval codes applied to simulated reflected light spectra of exoplanets \citep[see e.g.]{Lupu_2016, Lacy_2019}. \citet{Nayak_2017} find that observations at multiple phase angles can improve the constraints placed on atmospheric parameters such as radius, cloud top pressure and methane abundance for simulated observations of a cool gas giant planet based on the original planned WFIRST coronagraph configuration. \citet{Damiano_2020} demonstrate that extending the spectrum to longer wavelengths (as could potentially be the case for HWO) also breaks the degeneracy between parameters. Retrieval simulations so far demonstrate that gas abundances can indeed be constrained independently of cloud properties for cool Jupiter exoplanets in reflected light. 

\begin{figure*}[ht!]
    \centering
    \includegraphics[width=0.4\textwidth]{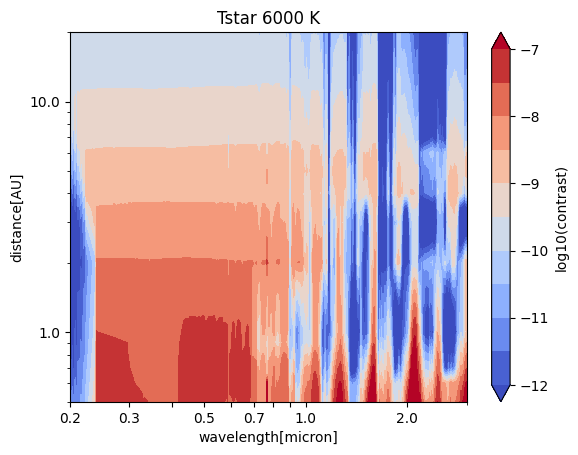}
    \includegraphics[width=0.4\textwidth]{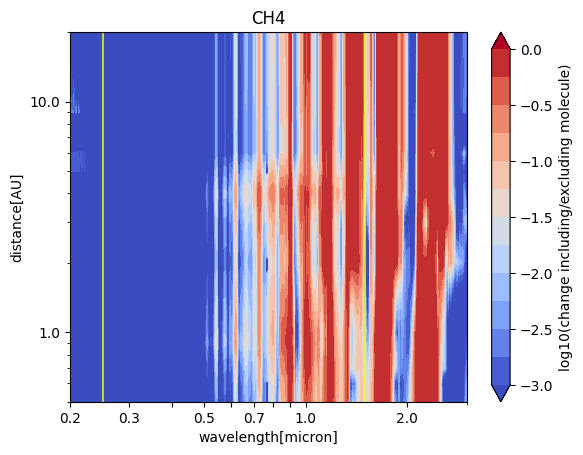}
    \includegraphics[width=0.4\textwidth]{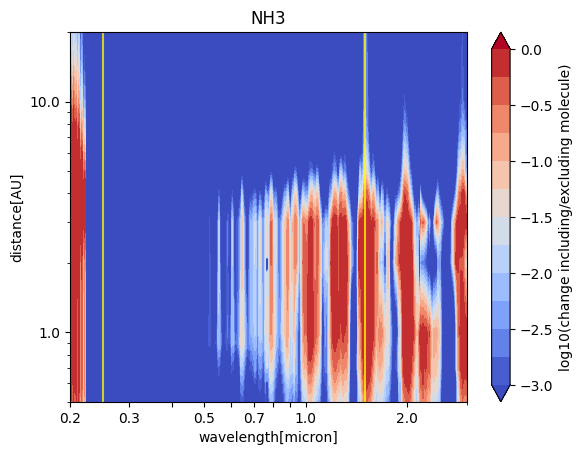}
    \includegraphics[width=0.4\textwidth]{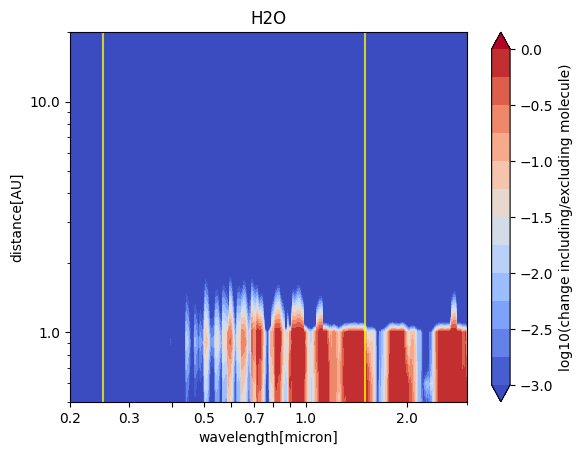}
    \caption{Simulated contrast as a function of wavelength for a Jupiter sized planet around a 6000K host star at different distances. The upper left panel shows the contrast of the planet compared to the host star. The other panels display the difference of the spectrum when given species are removed from the atmosphere. This can be used to estimate the SNR needed to detect a certain molecule at a given wavelength (e.g. a contrast of 0.1 would require a SNR of 10 to be detected).}
    \label{fig:contrast}
\end{figure*}

\begin{figure}[ht!]
    \centering
    \includegraphics[width=0.4\textwidth]{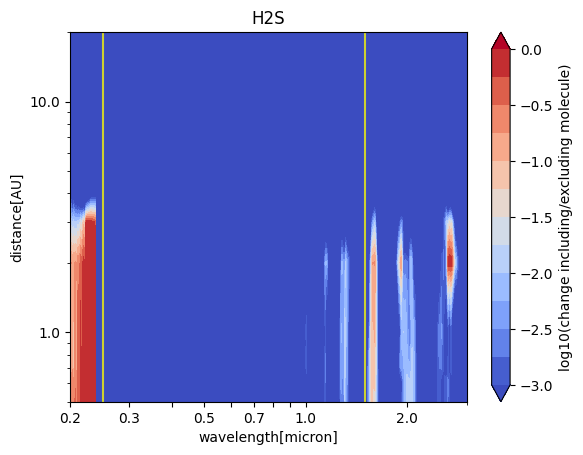}
    \caption{As Fig.~\ref{fig:contrast} but for the molecule H$_2$S.}
    \label{fig:contrastH2S}
\end{figure}

In Fig.~\ref{fig:contrast} we display the contrast of various simulated planets and the contrast of various important molecules in the atmosphere. This contrast is defined as the difference between the simulated spectrum including and excluding the given molecule. Roughly speaking a contrast in this plot of 0.1 would require a signal to noise of 10 to see the feature (as the feature is ten times smaller than the signal). When increasing the distance of the planet from its host star, the \add{gas phase composition of the} atmosphere transforms from H$_2$O dominated to NH$_3$ dominated and finally to CH$_4$ dominated. This is caused by the temperature difference and dominated by the rainout of various atoms: first water clouds, and for the coldest planets ammonia clouds. Note that for some distances to the central star there are some non-continuous features as a function of orbital distance. These are caused by a non-convergence of the cloud structure computations and reflect the complexity of this problem for intermediate cases (i.e. where a certain species is not yet fully rained out). From Solar System observations we do know that Saturn has a decreased abundance of NH$_3$ compared to Jupiter which is attributed to rainout of nitrogen because Saturn is significantly colder. Identifying at which temperatures the depletion of nitrogen starts will put strong constraints on cloud formation models. The yellow vertical lines in the contrast plot correspond to the nominal lower and upper wavelength cutoffs of 0.25 and 1.5\,$\mu$m respectively. It can be seen that for these molecules, this provides sufficient coverage of spectral signature. Ammonia has some absorption around 0.2\,$\mu$m which could be traced in principle when a lower wavelength cutoff is considered, but there are enough spectral features to constrain this molecule at longer wavelengths. A notable molecule that would require a slightly longer wavelength cutoff is H$_2$S, which has not been observed in gas phase on Saturn or Jupiter but is seen in the atmospheres of Uranus \citep{Irwin_2018} and Neptune \citep{Irwin_2019}. In Fig.~\ref{fig:contrastH2S} we show the contrast plot for H$_2$S from our simulated planets. It can be seen that the dominating feature visible for H$_2$S is located at 1.6\,$\mu$m, so a longer wavelength cutoff of 1.7\,$\mu$m or a short wavelength cutoff shorter than 0.25\,$\mu$m is required to capture this molecule. The 1.6\,$\mu$m feature is quite weak, so further study is needed to determine if the presence or absence of H$_2$S could be derived from spectra up to 1.7\,$\mu$m. \add{Additionally, H$_2$S could be constrained in the UV as can be seen from the strong signal below $0.25\mu$m in Fig.~\ref{fig:contrastH2S}.} Sulfur is expected to rain out in the form of NH$_4$SH (like in Jupiter) around 1 AU from a solar type star (see also Fig.~\ref{fig:contrastH2S}).

\section{Physical Parameters}

For the planets we need to measure:
\begin{itemize}
    \item The abundances of key molecules such as H$_2$O, CH$_4$, NH$_3$, PH$_3$ and H$_2$S, \add{C$_2$H$_2$} and potentially isotopologues such as CH$_3$D
    \item The characteristics of aerosols, including the composition, particle size and optical depth
    \item The upper atmosphere temperature structure
\end{itemize}

A key challenge that will need to be addressed for modelling these targets is that, unless the target is transiting, the planetary radius will not be known and will need to be recovered from the observations simultaneously with the above properties. Whilst some mass measurements may be available for objects with GAIA data, this is also likely to be unknown and may be another source of degeneracy in retrievals. 
For targets where such measurements are plausible (i.e. planets in close to edge-on orbits) we also aim to measure:
\begin{itemize}
    \item Phase-resolved atomic and molecular abundances (dayside, terminator/limbs, differences between evening \& morning limbs, and nightside)
    \item Phase-resolved cloud maps including cloud condensation pressure (and thereby temperature constraints)
    \item Longitudinally-resolved albedo across multiple rotation periods
    \item Rotation period of the planet
\end{itemize}
\add{These phase-resolved observations put constraints on stability of potential instrument systematics with time and star-planet separation to be able to properly assess if changes are due to true planet variation.}

\subsection{Atomic and molecular abundances}

\begin{figure*}[ht!]
    \centering
    \includegraphics[width=0.75\textwidth]{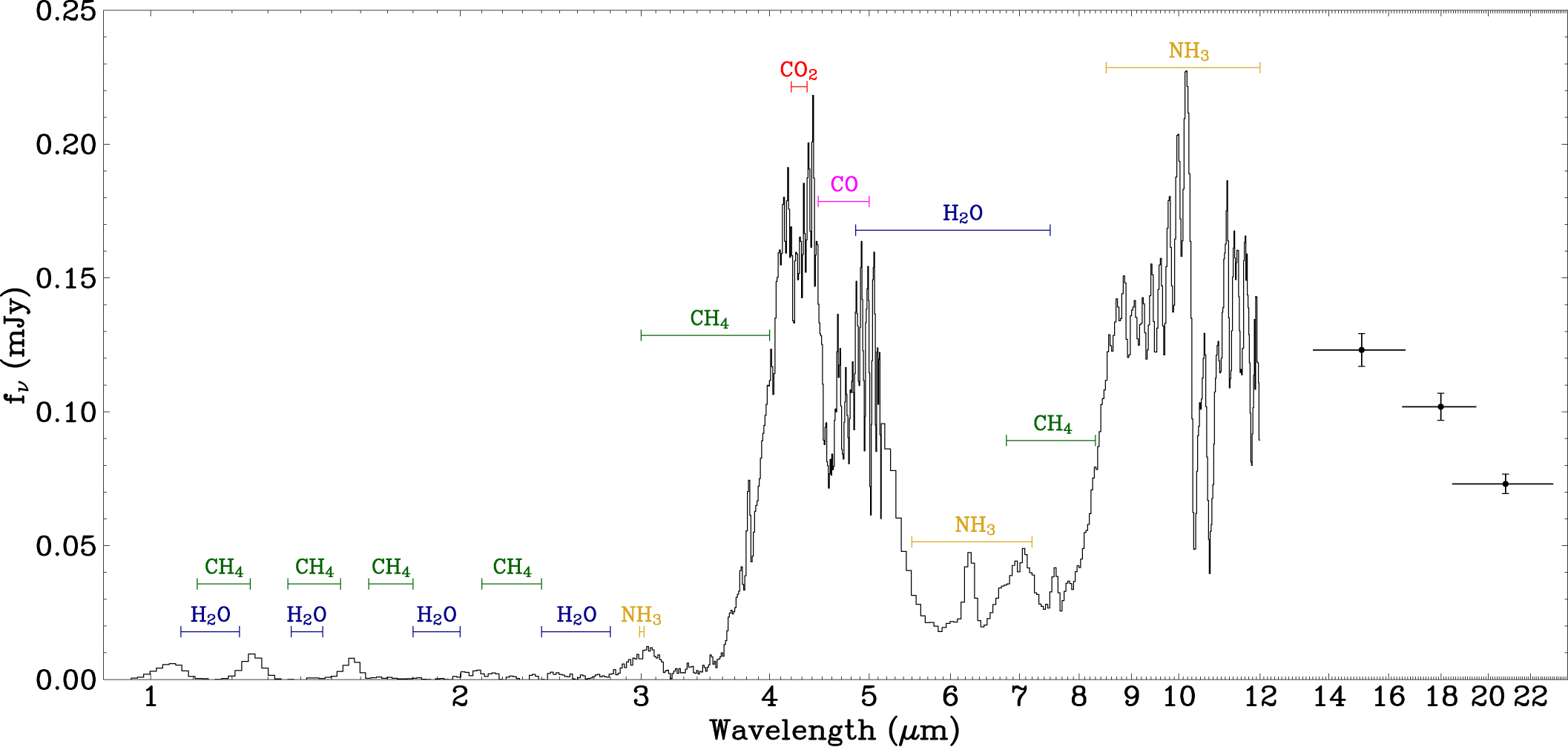}
    \caption{Taken from fig 2 of \citet{Beiler_2023} (used with permission), showing the first full spectral energy distribution from 1–21 $\mu$m of a Y-dwarf (WISE 0359-54). Absorption features due to many of the species we expect to occur in cool gas giants are present.}
    \label{fig:Beiler2023}
\end{figure*}

\emph{State of the art:} we do not yet have these measurements for any cool gas giants apart from those in the Solar System, with a very few exceptions such as Epsilon Indi Ab \citep{Matthews_2024} which has photometric data from JWST. The closest analog measurement with current capabilities is spectra of the cool brown dwarfs, which can be obtained using JWST (see e.g. Fig.~\ref{fig:Beiler2023}). This is not a perfect analogy since Y dwarfs are generally isolated objects (therefore do not require a coronagraph) and JWST is sensitive to their infrared thermal emission rather than optical wavelengths, but many of the molecules we hope to observe are the same. Fig.~\ref{fig:Beiler2023} shows absorption in a ~450K Y dwarf, with features due to H$_2$O, CH$_4$ and NH$_3$. Low resolution spectroscopy with R~100 is sufficient to capture the molecular features of the dominant molecules in the atmosphere, but would not allow the measurements of trace species e.g. isotopologues.

In the 2040’s it is expected that the ELTs will have identified several molecules in giant planets mostly around cooler stars. For the wide separation planets around Solar type stars the limiting contrast from the ground of $\sim 10^{-8}$ to maximum $10^{-9}$ limits the characterization to relatively close-in planets (more details discussed below). 

\emph{Required observations:} we will require coronagraphic observations of planets in orbits $>\sim0.5$\,AU from their parent stars in order to cover the transition from objects without water clouds to water cloud objects. We require spectroscopy covering the UV-optical-near IR range. See requirements section for details of wavelength ranges.  For phase-resolved measurements, we will need to make observations of the same target with appropriate time intervals for the phase to change, assuming the target’s orbit is favorable for this. 

\emph{Sensitivity/observational requirements:} The location of the near-infrared cutoff is likely to have substantial implications for observations of these targets, as is the contrast we are able to achieve. Extending the near-infrared coverage to $\sim5$\,$\mu$m will potentially allow us to observe auroral emission from these planets (the H$_3^+$ feature) and will provide us with multiple bands of NH$_3$, PH$_3$ and additional molecules (see Fig.~\ref{fig:Fletcher2023}). Observing across a wider wavelength range will also allow sensitivity to absorption at different pressure levels within the atmosphere, which will improve ability to characterize the atmospheric structure.

Realistically, extending the wavelength range to at least 1.7\,$\mu$m allows to capture one spectral band of the H$_2$S molecule which might allow some constraints on the dominant sulfur bearing species.
Contrast is critical for all observations detailed here. The LUVOIR final report details the expected contrast ratios for various gas giant planets, with a 2\,AU Jupiter reaching contrasts of order $10^{-8}$ and Jupiter at 5\,AU order $10^{-9}$. This extremely high contrast imaging will require excellent coronagraphic correction of the star and highly sensitive detectors. 

Although R$>1000$ spectroscopy has proven to be an important part of atmospheric characterization for known exoplanets (from hot Jupiters to directly-imaged planets), the benefits and science cases of moderate to high-resolution spectroscopy for reflected light planets remain an open question. It may increase the robustness of composition estimates, detect trace molecules such as isotopologues, measure radial velocities and spin-rotation (orbital parameters, spin-orbit alignment, etc.), or even resolve 2D surface features through Doppler imaging with high enough S/N and spectral resolution. Because the planetary spectral features are distinct from the star, higher spectral resolution can be significantly more robust to residual speckle noise. We therefore highlight the need for detailed high-resolution models and retrievals to explore the spectral resolution trade space and associated science cases. 

\subsection{Characteristics of aerosols}

\emph{State of the art:} as with the molecular observations above, we have not yet made observations of the type of target we propose here. Measurements of Solar System objects provide the best indicators of how this will proceed. Cloud properties of Solar System giants are generally determined through reflected light spectroscopy \citep[see e.g.][]{Irwin_2024} and also polarimetric measurements \citep[see e.g.][]{McLean_2017}. For Solar System objects, this allows spatial mapping of the cloud as the planets are resolved. 
In the 2040’s we might have a few observations of photometric polarization measurements from the Nancy Grace Roman Space Telescope (NGRST). This will be only possible for the closest stars and favorable planet orbits. Similarly, the ELT’s could reach the planets at smaller orbits and with relatively large contrast from the ground.

\emph{Required observations:} as above, we will require coronagraphic spectral observations of a range of planets. The optical wavelength region is likely to be most informative about cloud properties but their influence will also extend into the infrared. Measurements of linear polarization across a sample of planets would also be informative. As above, for cloud mapping purposes for suitable targets we would require these measurements to be made at a range of phases. 

\emph{Sensitivity/observational requirements:} many of the requirements are similar to those for molecular spectroscopy above; however, polarized light measurements would be a useful addition for this topic. Being able to measure the degree of linear polarization as a function of phase would be especially informative. 

\subsection{Upper atmosphere temperature structure}

\emph{State of the art:} Once again, the state of the art is largely limited to the Solar System. Constraining the temperature structure from reflected light measurements is actually relatively challenging, and this is inferred from the molecular absorption features we discuss above. Our ability to infer temperatures is enhanced by increased precision and spectral resolution in molecular bands, since the shape of the absorption feature is dependent on temperature but changes can be subtle. 

\add{Auroral processes may also play an important role in shaping the upper atmosphere and aerosol properties of giant exoplanets. Brown dwarfs and directly imaged companions are expected to have magnetic fields, and UV auroral emission has been predicted at levels potentially detectable by HWO \citep[e.g.]{Saur_2021}. Recent studies suggest that aurorae can influence the thermal structure \citep{Faherty_2024, ODonoghue_2021}. Access to UV auroral tracers, in combination with optical and IR spectra, would offer new insight into the coupling between magnetic activity and atmospheric chemistry in cool giant planets.}

In the 2040’s the ELTs will have put some constraints on the upper atmosphere structure of bright giant planets \add{at close-in orbits. For the planets at wider separation, ground based observation cannot reach these objects because of the contrast limit due to atmospheric seeing}.

\emph{Required observations:} as above, coronagraphic spectral measurements need to be made. To constrain temperature structure we must also obtain constraints on molecular species. 

\emph{Sensitivity/observational requirements:} these are also as above. 

\subsection{Longitudinally-resolved albedo across multiple rotation periods}

\emph{State of the art:} The state of the art is largely limited to the Solar System planets and massive brown dwarfs which are much more massive and hotter than cold gas giant exoplanets. In the 2040’s also here, the ground based ELT’s might have taken observations for some bright planets at large separation.

\emph{Required observations:} Photometric or low-resolution spectroscopic monitoring of a giant planet across multiple rotation periods (with coronagraphy). 

\emph{Sensitivity/observational requirements:} Accurate photometry or spectroscopy needs to be achieved within a fraction of a rotation period to derive longitudinally-resolved maps. For a giant planet with a similar rotation period as Jupiter (~10 hours), a signal-to-noise of 10 in 1 hour of observations is needed.

\subsection{Sample/survey size}
\label{sec:sample}

Cool gas giants in other planetary systems are a completely unexplored population, inaccessible to current observational techniques. These planets fill in an important gap in parameter space linking other planetary systems with our own Solar System, and observing them is something which, for Solar type stars, we will only be able to achieve with HWO. For cooler stars, there is a clear synergy with the second generation of ELT instrumentation which will be able to achieve contrasts and spatial resolutions suitable to image part of the parameter space. A population study of atmospheric composition and aerosol properties will be critical data for our understanding of the formation and evolution of gas giant planets at irradiation temperatures below 400 K (i.e. for Solar type stars distance from the central star $>0.5$\,AU). \add{To enable robust comparative planetology across stellar types, it will be important to ensure that the final target list includes a sufficient number of nearby M-dwarfs.}

\begin{figure*}[ht!]
    \centering
    \includegraphics[width=0.45\textwidth]{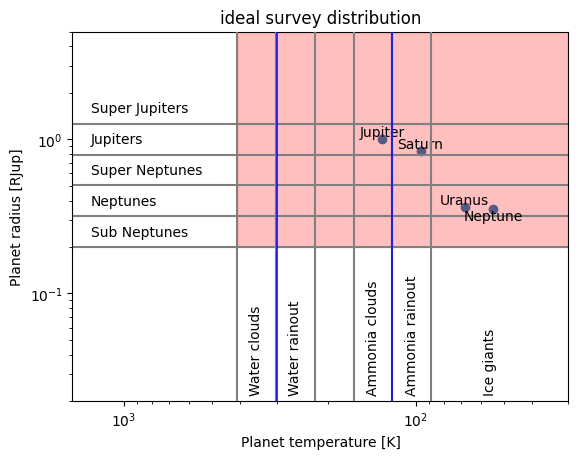}
    \includegraphics[width=0.45\textwidth]{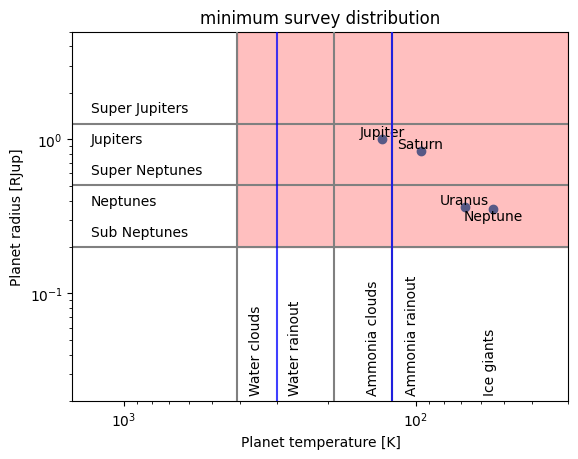}
    \caption{Proposed sample for comparative planetology for gaseous planets. The sample covers water and ammonia condensation/evaporation at various planet sizes.}
    \label{fig:survey}
\end{figure*}

\add{One subset of the gas giant population that HWO has exceptional potential to advance our understanding is the sub-Saturn class (labelled as super Neptunes in Fig~\ref{fig:survey}). These planets occupy the mass–radius space between Neptune and Saturn ($20\,M_{\oplus}\lesssim M_p<100\,M_{\oplus};\,4\,R_{\oplus}\lesssim R_p<8\,R_{\oplus}$) and have no analogue in the Solar System, leaving it unclear whether their compositions more closely resemble Neptunian or Jovian worlds. Density measurements indicate that sub-Saturns possess substantial envelope fractions, with 10–50\% of their total mass residing in their atmospheres~\citep[e.g.][]{2014_Lopez, 2018_Petigura, 2020_Petigura}. Although most targets reside within the Neptune desert, ridge and savanna, detections to date have remained sparse~\citep[e.g.][]{2024_Hacker, 2025_Baliwal, 2025_Thomas}. Of the $\sim200$ confirmed targets, about one-third do not transit their host stars, and among those that do, nearly 60\% have orbital periods longer than 10 days. Consequently, most of this population has remained either challenging or beyond the reach for atmospheric characterization with current instrumentation. The capabilities of HWO will enable an opportunity to observe across a much larger subset of sub-Saturns, refining our understanding of the transition between the upper limit of Neptunian worlds from the onset of Jovian planet formation.}

Obtaining phase-resolved data over multiple orbits is important because we want to understand the true composition of the planet in the presence of longitudinal temperature variations and disequilibrium chemistry and be able to remove signals due to stellar variability. Whilst these measurements will not be achievable for all targets, it is crucial to perform them for a significant subset in order to understand the phase dependence of our measurements across the full sample of targets. \add{Observations of close to full day and close to full nightside of the planet will be most challenging because of the small projected separation. Also, the nightside emission will not be observable for these targets as the nights do not scatter light and thermal emission will be faint.}

Fig.~\ref{fig:survey} sketches the proposed survey sample ranging over planet radii and temperatures. For a proper comparative planetology study we need at least one planet per box in the rightmost panel of Fig.~\ref{fig:survey} allowing us to study the influence of planet size and temperature (i.e. 12 boxes/planets at the minimum). For a more robust picture of transitions between planet sizes and temperature regimes the left panel represents a robust survey size. There are 30 boxes to fill, leading to the rough estimate that at least 30 planets need to be found available for robust characterization of trends. Even more ideally, we fill this parameter space for Solar type stars and for M-dwarfs, roughly doubling the required sample. Here it should be noted that filling the lower right part of the plot puts the firmest constraints on contrast at the outer working angle. In addition, synergy with next generation ELT instrumentation is likely required to fill the entire parameter space for Solar type and M-dwarf stars.

\subsection{Comparison with ELT sample and 2040’s state of the art}
\label{sec:compareELT}

\begin{figure*}[ht!]
    \centering
    \includegraphics[width=0.45\textwidth]{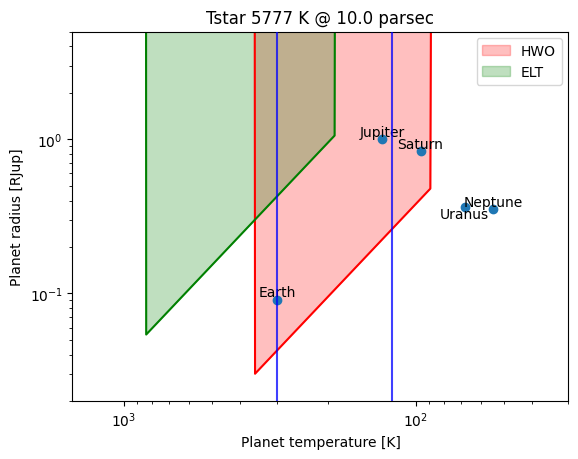}
    \includegraphics[width=0.45\textwidth]{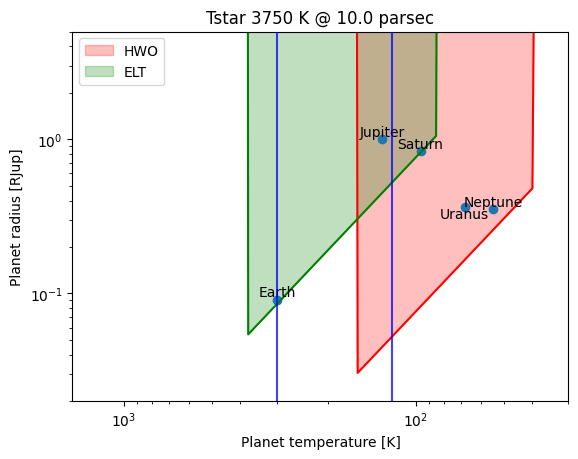}
    \caption{Comparison of the planets within reach for a 30m ground based ELT and a 6.5m space based HWO for stars at 10 parsec. The Solar System Giants (and Earth) are plotted in there for comparison. The blue vertical lines represent the expected temperatures where H$_2$O and NH$_3$ will rain out.}
    \label{fig:contrastHWO_ELT}
\end{figure*}

At the time when HWO will fly, we will have ground based ELTs online with extremely accurate high contrast imaging capabilities. To compare the sample that these telescopes will be able to provide with the capabilities of HWO, we follow the following exercise. We compute as a function of planetary effective temperature (which translates to orbital distance given a certain central star) the smallest possible planet still detectable with a given telescope. For the ELTs we take for this the maximum contrast of $10^{-8}$. The ELTs are expected to reach close to diffraction limit in spatial scale by using adaptive optics, but will be limited in contrast due to atmospheric conditions. Note that these parameters are indicative for the second (or higher) generation instrumentation for the ELTs and are not likely to be achieved at first light. For HWO we take the contrast to be $10^{-10}$. For the inner- and outer working angle we take 3 to 40 lambda/D as the range available for the high contrast imaging system. We take a 30m ELT and a 6.5m HWO configuration. We compute the planets within reach given a star at 10 parsec and plot them in Fig.~\ref{fig:contrastHWO_ELT}. As can be seen, for Solar type stars planets similar to Jupiter and Saturn are in reach for HWO, but due to the low contrast this cannot be reached by the ELT. In addition, for Solar type stars any planet larger than Neptune and warmer than Saturn would be within reach of HWO. For planets around M dwarf stars with similar temperatures, we see that the Jupiter and Saturn type planets are within reach of both the ELT and HWO. We conclude that for Solar type stars the ELTs will likely image a large fraction of the hot to warm Jupiters, while HWO will be ideal for the cooler gas Giants. For M dwarfs, the ELTs will be able to already image some of the cool gas Giants, while HWO has access to the coldest ice Giants similar to Uranus and Neptune in temperature.

\begin{table*}[tbh!]
    \centering
    \caption{Physical parameters to be constrained.}
    \label{tab:parameters}
    \begin{tabularx}{\textwidth}{X||X|X|X|X}
        \noalign{\smallskip}
        \noalign{\smallskip}
Physical Parameter &
State of the Art &
Incremental Progress (Enhancing) &
Substantial Progress (Enabling) &
Major Progress (Breakthrough) \\
        \hline
        \hline
Number of Objects &
Roughly 10, over lifetime of JWST, (potentially several with NGRST), and several dozen with ELT &
12 over a broader parameter range &
30 with some of them in single systems &
$>$60 in systems orbiting different types of stars \\ 
        \hline 
Planet Temperature Range/Distribution &
Ranging from 200-2000K (Temperature from formation) and Solar System planets &
Ranging from 100K to 400K effective temperature &
Ranging from 30K to 600K effective temperature \\
        \hline 
Host Star Temperature Range/Distribution &
Mostly F-K stars for hot Jupiters, mostly M type for colder planets &
 F-K type stars for a broader range of planet temperatures &
 F-M \\
        \hline
    \end{tabularx}
\end{table*}

\subsection{\add{Expected statistics from blind search}}
\label{sec:yield}

\begin{figure*}[ht!]
    \centering
    \includegraphics[width=0.32\textwidth]{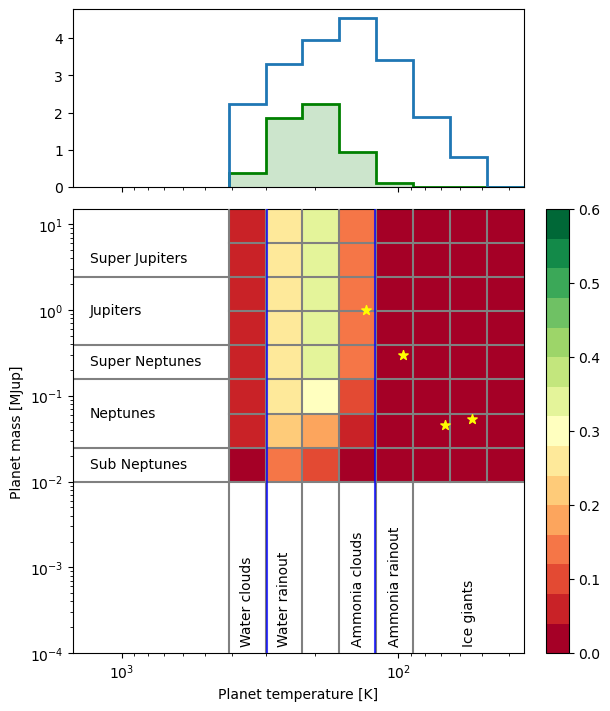}
    \includegraphics[width=0.32\textwidth]{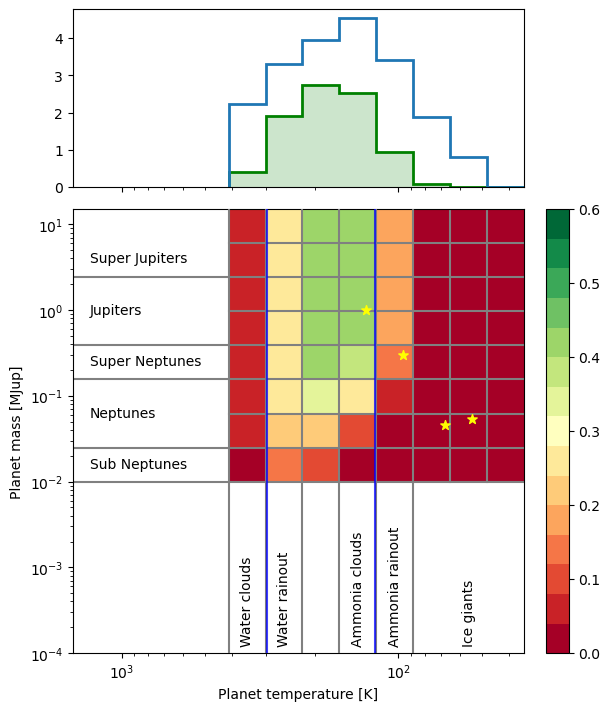}
    \includegraphics[width=0.32\textwidth]{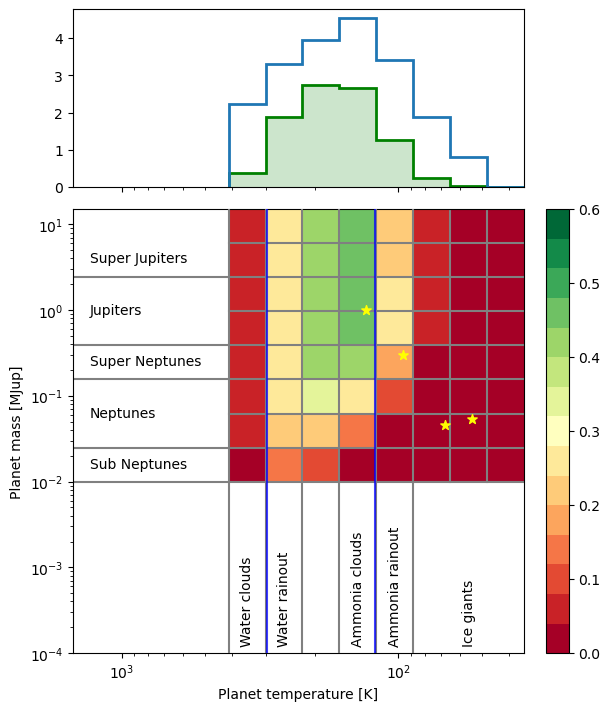}
    \caption{\add{Expected statistics of gaseous planets from a blind survey of 164 targets selected to be promising habitable exoplanet host stars for HWO. The left figure shows the case for an outer working angle (OWA) of $10\lambda/D$, the middle one for an OWA of $25\lambda/D$ and the right one for $40\lambda/D$. The top panels show the distribution of detectable planets as a function of temperature in green and the underlying true population in blue. The yellow stars indicate the location of Jupiter, Saturn, Neptune and Uranus in this plot.}}
    \label{fig:StatisticsOWA}
\end{figure*}

\add{The occurrence rate of cold gas giant planets is currently still unknown because of biasses in the detection methods. However, estimates do exist and allow us to make an educated guess on the expected yield of cold gaseous planets from a blind search of nearby stars. For this we use the same setup as in the previous section for the telescope detection limit. We use the statistics from \cite{Fulton_2021}. This gives roughly 14 cold gaseous planets per 100 stars. As input sample we use the catalogue from \citet{Harada_2024}. This catalogue contains 164 promising targets for HWO to survey for habitable planets. We compute what the statistics would be for randomly catching gaseous planets while surveying the stars in this catalogue. The yield is computed by randomly creating planets around the sample stars and predicting their brightness and planet-star separation using a simple albedo model with a Lambertian phase curve. The results for the blind survey for various outer working angles is shown in Fig.~\ref{fig:StatisticsOWA}. This is for a wavelength of $0.75\,\mu$m and a limiting contrast of $10^{-10}$.}

\begin{figure*}[ht!]
    \centering
    \includegraphics[width=0.32\textwidth]{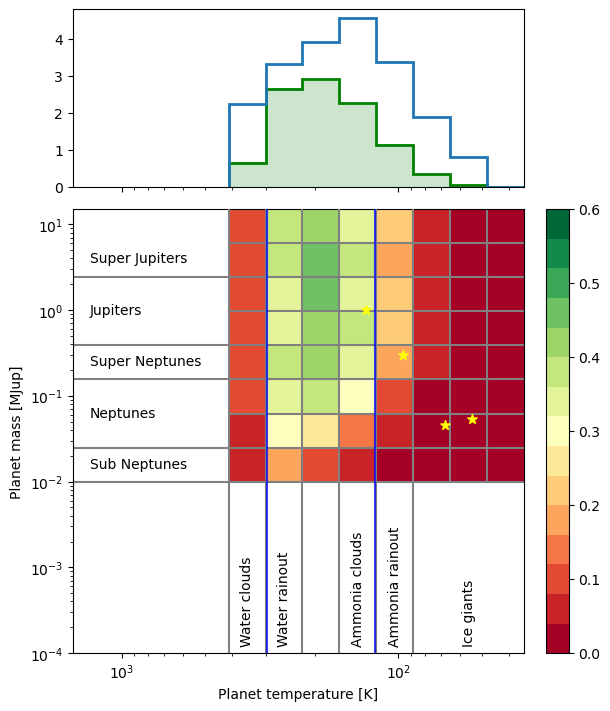}
    \includegraphics[width=0.32\textwidth]{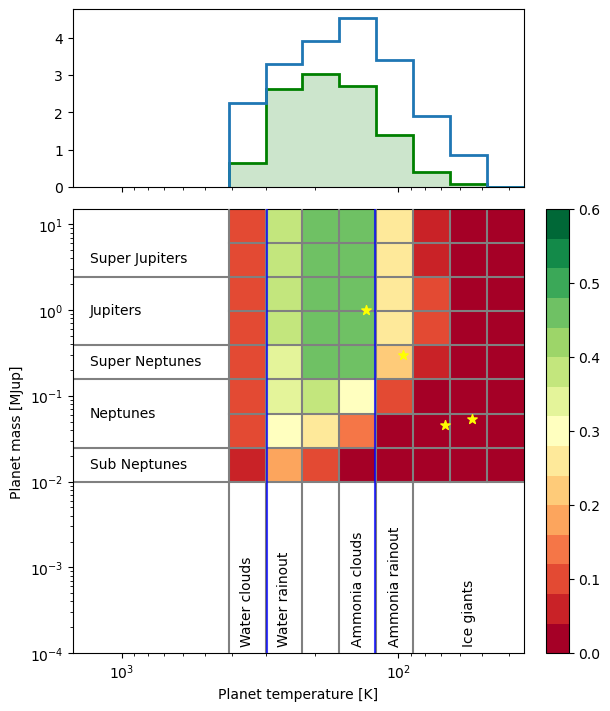}
    \includegraphics[width=0.32\textwidth]{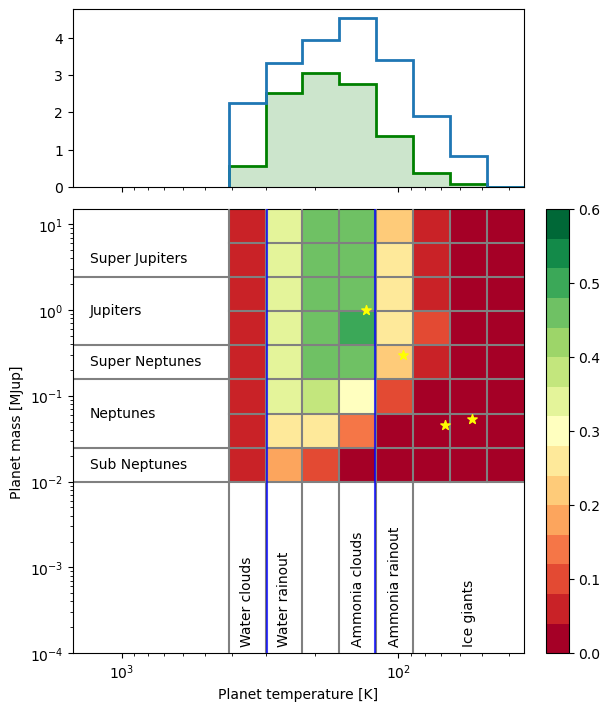}
    \caption{\add{Same as Fig.~\ref{fig:StatisticsOWA} but for the timing of observations optimized to capture known planets in the system.}}
    \label{fig:StatisticsOpt}
\end{figure*}

\add{It is expected that from radial velocity measurements the orbital parameters of several gian planets will be known by the time that HWO will observe. Assuming we have this data for all 164 stars in the sample, we can optimize the timing to make sure the planets are observable. If we do this, we can significantly increase the yield. This is shown in Fig.~\ref{fig:StatisticsOpt}.}

\begin{figure*}[ht!]
    \centering
    \includegraphics[width=0.32\textwidth]{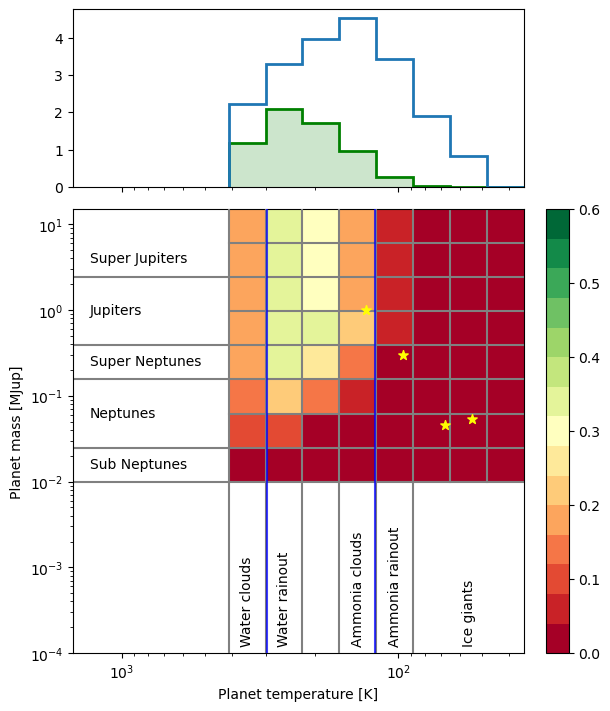}
    \includegraphics[width=0.32\textwidth]{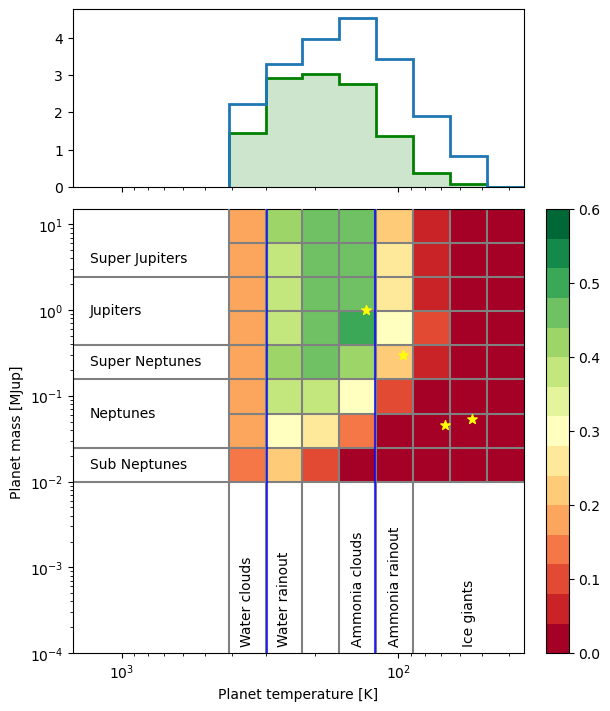}
    \includegraphics[width=0.32\textwidth]{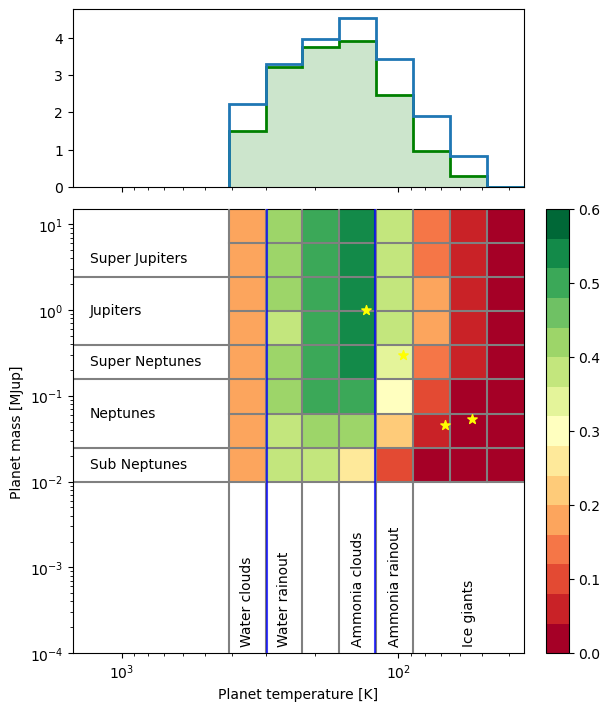}
    \caption{\add{Same as Fig.~\ref{fig:StatisticsOWA} for different values of the limiting contrast. All panels are assuming an OWA of $20\lambda/D$, an optimal choice of the wavelength and optimized observation timing of known planets. The left panel is for a contrast limit of $10^{-9}$, the middle panel for $10^{-10}$ and the right panel for $10^{-11}$.}}
    \label{fig:StatisticsContrast}
\end{figure*}

\add{When the orbital parameters of a planet are known, we can also pick the wavelength of observation to match with the inner and outer working angle. For example, a planet that is known to orbit close to the star is best observed at small wavelengths (to decrease the IWA), while a planet orbiting at a larger distance is better observed at longer wavelengths (to increase the OWA). If we also optimize for that, we can increase the sample statistics. This is shown in Fig.~\ref{fig:StatisticsContrast} for three different values of the limiting contrast. As can be seen, if the limiting contrast is $10^{-11}$, a large fraction of planets are within reach of HWO capabilities as long as we optimize observational settings (wavelength and timing) also for an outer working angle of $20\lambda/D$.}

\add{Combining the statistics above, we conclude that the outer working angle is a critical factor in the number of cool gaseous planets accessible for HWO. An outer working angle of 20$\lambda/D$ corresponds to $\sim 1$arcsec at a wavelength of $\lambda=1.5\,\mu$m. This outer working angle is crucial for characterisation of planets colder than Saturn. If we assume we can optimize the observing time for optimal planet phase this implies that for capturing a signficant part of the NH$_3$-ice forming planets in the standard stellar input catalogue of HWO:
\begin{itemize}
    \item For cloud characterisation, which is best done at wavelengths $<\sim 0.7\,\mu$m, an outer working angle of 40$\lambda/D$ is required.
    \item For molecular characterisation, which is best done at wavelengths $>\sim 1\,\mu$m an outer working angle of 30$\lambda/D$ is required.
    \item For simple detection, which can be done at any wavelength, an outer working angle of 20$\lambda/D$ is required.
\end{itemize}
Note that a workaround on these constraints is to observe stars further away. The feasibility of this needs to be investigated given the brightness limits of the high contrast imaging mode of HWO.
}

\section{Description of Observations}

We will need to obtain UV, optical and near-infrared spectra with a coronagraph in place to block starlight for a range of gas giant planets at a range of orbital distances from their host stars around stars ranging from F dwarfs to K dwarfs. For targets in close to edge-on orbits, where we expect to see changes with phase, we will make multiple similar measurements throughout the course of an orbit, with the appropriate cadence for the planet’s orbital period (we would ideally target ~10 observations within a single period.) Ideally, measurements would be simultaneous across the entire wavelength range, but this is a less critical consideration for directly imaged planets compared with transit spectroscopy measurements as contamination from the parent star’s spectrum should be much less of a challenge. We will also obtain shorter cadence measurements for the most favorable targets where we can achieve the largest signal-to-noise, to look for variations in atmospheric properties over the planet’s rotation period; the cadence of these measurements will depend on the planet’s expected rotation period, which will probably be of order hours to days (considering Jupiter and Saturn have rotation periods of $\sim$10 hours and isolated brown dwarfs have rotation periods of $<$1 day), and therefore we will want to take measurements on the timescale of \add{5-10 hours to sample rotational modulation, ideally spanning a full rotation ($\sim10$ hours to 1 days) to cover at least one full rotation.} Measuring variability in this way will provide us with evidence of the planet’s rotation period.

The ideal spectral range based on our current understanding would be from 0.1 - 5\,$\mu$m, the spectral range shown for Jupiter in Fig.~\ref{fig:Fletcher2023}. This encompasses a range of spectral features from a variety of important gasses, including water vapor if present, methane, ammonia and phosphine. Measuring the abundances of these gasses will enable us to constrain the C-N-O-P elemental budget of the planet’s atmosphere; if sulfur species such as H$_2$S are present, we would also be able to place constraints on sulfur. The advantage of the 2 - 5\,$\mu$m range in particular is the access to more ammonia and phosphine bands, and also to H$_3^+$ emission, which if observed would indicate the presence of auroral activity. This would be extremely informative about the magnetic field properties of cool giant planets. The ultraviolet and optical portion of the spectrum, as well as being sensitive to molecular absorption, will also provide us with information about clouds and their scattering properties, and enable us to measure the albedo of the planet. Linear polarization measurements in the optical part of the spectrum, especially obtained at multiple phase angles for planets where this is possible, will further constrain the properties of aerosol particles that may be present. 

Requirements:
\begin{itemize}
    \item UV capability to measure reflected light and albedo 
    \subitem Here it is important to put more firm constraints on the need for how far into the UV would be required to constrain certain species.
    \item Time cadence to measure spatial changes or pixel resolution
    \subitem The exact cadance required depends on the expected rotation period for these types of planets. This is unknown, but the reference for the Solar System is hours to ~1day
    \item Polarization
    \subitem Ideally this includes spectro-polarimetry at various phase angles of the planet.
\end{itemize}

HWO will fills a unique and otherwise inaccessible part of the parameter space for UV-optical-IR spectroscopy of cool gaint planet atmospheres. There is a strong synergy with ground based observations using ELT second generation instrumentation capable of imaging the warmer, closer-in giant planets. Especially true Jupiter and Saturn analogs around Solar type stars are inaccessible for any current or future planned facility other than HWO \add{(with the exception on a few exceptionally favorable targets that might be accessible with the Roman Space telescope)}. Such planets are inaccessible to transit spectroscopy due to their relatively long orbital periods and low transit probability, and JWST is not able to reach the appropriate coronagraphic contrast to observe their thermal emission directly. HWO presents our first opportunity to properly study the atmospheres of true exo-Jupiters \add{and of exo-Neptunes around M-dwarfs}.

Main requirements are on the inner and outer working angle to ensure measurements of warm planets at small phase angles (inner working angle) and to ensure measurements of cool planets at quadrature phase. Here it is important to note that extending the outer working angle should be accompanied with increased contrast at the outer working angle as very cold planets far out are expected to be extremely dim (contrast < $10^{-10}$). The tradeoff on how the parameter space presented in Fig.~\ref{fig:survey} could be covered by various choices of telescope configurations should be conducted especially to judge the advantage of increasing the outer working angle. Enabling science is identified through polarimetric observations, and especially spectro-polarimetry, to robustly identify clouds and hazes.

\begin{table*}[tbh!]
    \centering
    \caption{Physical parameters to be constrained.}
    \label{tab:parameters}
    \begin{tabularx}{\textwidth}{X||X|X|X|X}
        \noalign{\smallskip}
        \noalign{\smallskip}
Observation Requirement &
State of the Art &
Incremental Progress (Enhancing) &
Substantial Progress (Enabling) &
Major Progress (Breakthrough) \\
        \hline
        \hline
Type (imaging, spectroscopy, etc.) &
Imaging of some types of system &
Imaging &
Imaging spectroscopy &
Imaging spectroscopy \\
\hline
Wavelength Range &
Likely some NIR will be done with the ELTs &
Visible &
0.2 - 1.8 micron &
0.15 - 5 micron \\
\hline
Inner working angle &
Possibly 150 mas (NGRST) &
0.25 arcsec (a 5AU Jupiter analog around a star 20 parsec away) &
$\sim$0.1arcsec (a Jupiter analog without H2O rainout around a Solar type star at 10 parsec) &
$\sim$60 mas (see Vaughan et al. 2024) which will allow scattering phenomena such as e.g. rainbows to be observed for a range of targets \\
\hline
Outer working angle &
Possibly 660 mas (NGRST) & \add{$20\lambda/D$: detection beyond the NH$_3$ ice line. Characterisation of clouds beyond the H$_2$O ice line. Characterisation of molecules up to and slightly beyond the NH$_3$ ice line.}& \add{$30\lambda/D$: characterisation of clouds up to the NH$_3$ ice line.} &
1 arcsecond \add{at all wavelengths} - observe a 10au planet at quadrature for a system 10 pc away \\
\hline
Polarization observations &
Possibly some with NGRST &
Photometric polarimetry &
Spectral polarimetry in some molecular bands &
Spectral polarimetry over the range of 0.2 - 1.8 micron \\
\hline
Amount of sky covered & & & &
This requires more detailed calculations\\
\hline
Magnitude of target in chosen bandpass & & & &
This requires more detailed calculations \\
        \hline
    \end{tabularx}
\end{table*}

\section{Identification of required follow up investigations}

\add{The following modeling and retrieval studies are necessary to refine the observational requirements for HWO’s characterization of cool gas giant exoplanets. These investigations should focus on the parameter space accessible to HWO as outlined in Fig. \ref{fig:contrastHWO_ELT} and Sections \ref{sec:sample}-\ref{sec:compareELT}. More specifically, planets with masses from Neptune to a few times Jupiter and orbital separations corresponding to irradiation temperatures below 400 K. This includes planets around both Solar-type and M-dwarf host stars, with a focus on cool Jupiters and sub-Jovians in temperate orbits.}
\begin{itemize}
    \item Investigate the required wavelength coverage, SNR and spectral resolution to obtain the required accuracies on the abundances in the atmosphere. Perform retrievals for varying planet properties and chemically expected abundances of H$_2$O, CH$_4$, NH$_3$ and potential other interesting species (list to be determined).
    \item Investigate the cloud properties that can be derived from retrieval. How do physical constraints on retrievals help here (constraints on irradiation temperature etc.)?
    \item Further investigate if the internal heat can be constrained like suggested in \citep{Hu_2019}.
    \item Further investigate what the needs are of additional constraints on planet mass and orbital radius for the retrievals to work.
    \item Determine sample statistics expected using the framework \add{section \ref{sec:yield} but with realistic contrast curves for the various telescope configurations and a broader range of input sample stars.}
\end{itemize}

{\bf Acknowledgements.} 

\bibliography{author.bib}

\appendix

\section{Comparison of model simulations with Jupiter}

\begin{figure*}[ht!]
    \centering
    \includegraphics[width=0.7\textwidth]{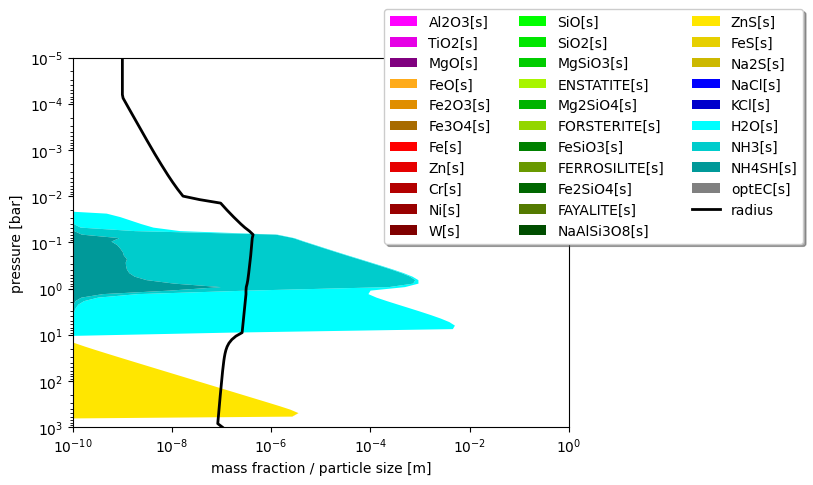}
    \caption{Cloud structure resulting from the modeling setup using the parameters of Jupiter around the Sun. As can be seen the upper cloud deck is very similar to the expected structure with NH$_3$, NH4SH and H$_2$O clouds. Deep down the model also predicts the formation of ZnS clouds.}
    \label{fig:clouds_jupiter}
\end{figure*}

The model simulations used in this paper were computed using the ARCiS modelling framework. Within this model we compute the temperature structure, chemistry and cloud formation self-consistently. For the cloud formation we use the method from \citet{Huang2024}. Fig.~\ref{fig:clouds_jupiter} shows the cloud structure predicted from the model. As can be seen the cloud layers formed by the cloud formation model are very comparable to the general consensus in the literature \citep[see e.g.][]{Atreya_2005}.

\begin{figure*}[ht!]
    \centering
    \includegraphics[width=0.45\textwidth]{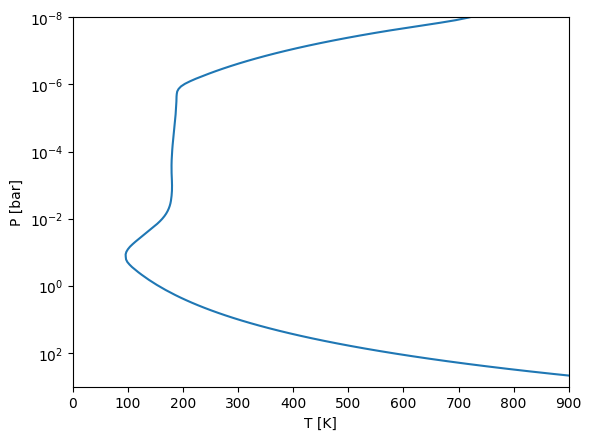}
    \caption{Temperature structure from the simulations using the Jupiter input values. The derived temperature structure from the self-consistent model is very similar to the observations from the Galileo probe.}
    \label{fig:PT_jupiter}
\end{figure*}

The predicted temperature structure for the Jupiter model is shown in Fig.~\ref{fig:PT_jupiter}. The overall predicted geometric albedo is 0.53 which is also very close to the observed value for Jupiter of 0.54.

\end{document}